\documentclass{appolb}
\usepackage{epsfig}

\begin{document}

\title{Coexistence of resonant activation and noise enhanced stability in a model of tumor-host interaction:
Statistics of extinction times}

\author{Anna Ochab--Marcinek, Ewa Gudowska--Nowak
\address{Marian~Smoluchowski Institute of Physics,
 Jagellonian University\\ and Mark Kac Complex Systems Research Center\\ Reymonta~4, 30--059~Krak\'ow, Poland,
 \footnote{ochab@th.if.uj.edu.pl}}
\and Alessandro Fiasconaro, Bernardo Spagnolo
\address{Dipartimento di Fisica e Tecnologie Relative and
INFM-CNR,\\
Group of Interdisciplinary Physics\footnote
{http://gip.dft.unipa.it}, Universit\`a di Palermo,\\ Viale
 delle Scienze, I-90128 Palermo, Italy}}

\maketitle

\date{\today}

\begin{abstract}
We study a Langevin equation derived from the Michaelis-Menten (MM)
phenomenological scheme for catalysis accompanying a spontaneous
replication of molecules, which may serve as a simple model of
cell-mediated immune surveillance against cancer. We examine how two
different and statistically independent sources of noise - dichotomous multiplicative noise and
additive Gaussian white noise - influence the population's extinction
time. This quantity is identified as the mean first passage time of
the system across the zero population state. We observe the effects
of resonant activation (RA) and noise-enhanced stability (NES) and
we report the evidence for competitive co-occurrence of both phenomena in a given
 regime of noise parameters. We discuss the statistics of first
passage times in this regime and the role of different pseudo-potential
profiles on the RA and NES phenomena.
 The RA/NES coexistence region  brings an interesting interpretation
 for the growth kinetics of cancer cells population, as the
NES effect enhancing the stability of the tumoral state becomes strongly
reduced by the RA phenomenon.

\end{abstract}



\section{Introduction}

Since the times of Delbr\"uck \cite{Del}, a considerable research has been done on auto-catalytic reactions
using stochastic models~\cite{Gillespie}-\cite{Paulsson}.
Spontaneous random molecular fluctuations play a significant role in
the driving kinetic mechanism of many regulatory enzymatic
reactions. Random fluctuations in genetic networks are inevitable as
chemical reactions are probabilistic and many genes, RNAs and
proteins are present in low numbers per cell. Gene expressions, for
example, involves a series of single-molecule events subject to
significant thermal fluctuations, which may result in a divergence
of fate and lead to non-genetic population heterogeneity. Moreover,
mean levels of heterogeneity for the populations are ensured by
regulatory factors that tune the spontaneous fluctuations. Although
many examples of potentially noise-exploiting cellular processes
have been discussed and documented in literature, how cells function
and process information when the underlying molecular events are
random still remains an open question
~\cite{Paulsson}-\cite{paulsson}. Other cellular processes
influenced by noise include ion-channel gating, neural firing,
cytoskeleton dynamics and molecular
motors~\cite{white}-\cite{simon}.

Simulation schemes that model each individual reaction event can be computationally demanding \cite{Gillespie} as the copy of the input and output molecules increases. Therefore, as an alternative way to account for molecular fluctuations in biochemical reactions is to assume that the time evolution of a species of interest is produced at an average rate $f(x)$, degraded at rate $g(x)$ and follows the stochastic differential equation:

  \begin{equation}
   \frac{dx}{dt}=f(x)-g(x)+\xi(t),
   \label{eq:lang}
 \end{equation}
In the above equation the noise term is appended as the additive (white) noise source $\xi(t)$ weakly perturbing deterministic evolution of the system. Obviously, implicit for such a formulation is  a continuous description of molecular species where the dynamics is cast in terms of infinitesimal changes of concentrations. 
Eq.~(\ref{eq:lang}) will be further postulated to study dynamics of
a catalytic reaction . We consider the phenomenological
Michaelis-Menten scheme for the catalysis accompanying a spontaneous
replication of molecules
   \begin{eqnarray}
  X &\longrightarrow& \!\!\!\!\!\!\!\!\!\!^{\lambda} \quad 2X  \nonumber\\
 X\ +\ Y\  &\longrightarrow&
 \!\!\!\!\!\!\!\!\!\! ^{k_{1}}\ \ \ \ Z\ \longrightarrow
 \!\!\!\!\!\!\!\!\! ^{k_{2}}\ \ \ \  Y\ +\ P\ ,
 \label{scheme}
 \end{eqnarray}
where a substrate $X$ forms first a complex $Z$ with molecules of
the enzyme $Y$, before the conversion of $X$ to a product $P$ is
completed. By assuming that the production of $X$-type molecules
inhibited by a hyperbolic activation (Michaelis-Menten inhibition kinetics) is the slowest process under consideration and by considering a conserved mass of enzymes
$Y+Z=E=const$, the resulting kinetics can be recast in the form of
the 1-dim Langevin equation

\begin{equation}
\frac{dx}{dt}=-\frac{dU(x)}{dx}+\xi(t)
\label{lang}
\end{equation}
with the pseudo-``free-energy" potential profile $U(x)$ expressed
as~\cite{Prigogine}
 \begin{equation}
  \label{eq:pot}
  U(x)=-\frac{x^2}{2}+\frac{\theta x^3}{3}+\beta x -\beta
  \ln(x+1),
  \label{profile}
 \end{equation}
where $x$ is the normalized molecular density with respect to the
maximum number of molecules, and with the following scaling
relations
 \begin{equation}
 x=\frac{k_{1}x}{k_{2}}, \,\,\,\,\,   \theta=\frac{k_{2}}{k_{1}}, \,\,\,\,\,
\beta=\frac{k_{1}}{\lambda}, \,\,\,\,\, t=\lambda t.
\label{eq:scaling}
 \end{equation}

The time-scale separation performed on a kinetic scheme of the system (\ref{scheme}) requires that a subset of species is asymptotically at steady state on the time scale of interest, and is known as a quasi-steady-state assumption \cite{Paulsson,rao}. It reduces the model complexity by effectively reducing the number of considered molecular species and reactions.\\

\begin{figure}[t]
\begin{center}
\epsfig{figure=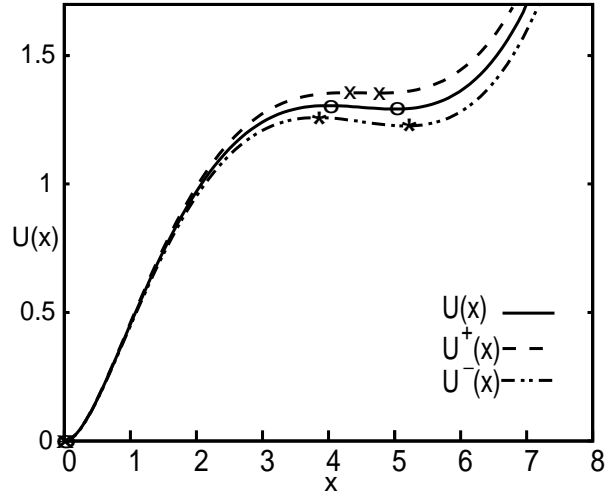, width=8cm}
\end{center}
\caption{\label{fig:pot} The Michaelis-Menten potential with
parameters: $\beta=3, \theta=0.1, \Delta=0.02$. Labels: ``o":
extrema of $U(x)$:$x = 0, 4, 5$; ``x": extrema of $U^+(x)$: $x=0,
4.28, 4.72$; ``*": extrema of $U^-(x)$: $x=0, 3.83, 5.17$.
 }
\end{figure}

 In the case of Michaelis-Menten kinetics this approximation assumes also much larger concentration of
substrate $X$ than the enzyme concentration $E$ and with the steady-state constraint $dZ/dt=0$ leads to a characteristic regulatory term $-\beta x/(1+x)$ in the dynamic (\ref{lang}).
The resulting potential profile eq.(\ref{profile}) has at most three extrema representing
deterministic stationary states of the system (see
Fig.~\ref{fig:pot})
\begin{eqnarray}
 x_1&=&0, \\
 x_2&=&\frac{1-\theta+\sqrt{(1+\theta)^{2}-4\beta
   \theta}}{2\theta}, \\
 x_3&=&\frac{1-\theta-\sqrt{(1+\theta)^{2}-4\beta
   \theta}}{2\theta}.
 \end{eqnarray}

The essential feature captured by the model is, for a constant
parameter $\theta$, the $\beta$-dependent bistability. In the above
form and by assuming time dependent, random variations of the
parameter $\beta$, the model has been used to describe  an effect of
cell-mediated immune surveillance  against the cancer \cite{GARAY}.
 Most of tumoral cells bear antigens which are
recognized as strange by the immune system. A response against these
antigens may be mediated either by immune cells such as
T-lymphocytes or other cells, not directly related to the immune
system (like macrophages or natural killer cells). The process of
damage to tumor proceeds via infiltration of the latter by the
specialized cells which subsequently develop a cytotoxic activity
against the cancer cell-population. The series of cytotoxic
reactions between the cytotoxic cells and the tumor tissue may be
considered to be well approximated by a saturating, enzymatic-like
process whose time evolution equations are similar to the standard
Michaelis-Menten  kinetics. The variability of kinetic parameters
defining this process  naturally affects the extinction of the tumor
\cite{GARAY,lefever}.

The process of population growth and decay can be described as a
motion of a fictitious particle in a potential $U(x)$, whose shape
varies in time due the noisy $\beta$ parameter. Minima of the
potential correspond to stable states of the population, where, in
case of a purely deterministic system, the cell concentration does
not grow or decay. One of the minima is always at $x=0$, which
reflects the fact that the state in which cancer cells are absent is
a stable one. The other minimum corresponds to a fixed-size cell
population. Noise-induced transition between those stable states
across the potential barrier may be interpreted as a "spontaneous"
extinction of cancer due to environmental fluctuations, as well as
due to random variations in the immune response efficiency. We will
study a system with parameters chosen in such a way that $x=0$ is a
global minimum. We are interested in the lifetime of the metastable
state. Mean escape time has been intensively studied in order to
characterize the lifetime of metastable  states of static and
fluctuating potentials with different initial conditions
\cite{doe}-\cite{ale}. The studies show that the mean escape time
has different non-monotonic behaviors as a function of both the
thermal noise intensity and the mean frequency of potential
fluctuations. These behaviors are a signature of two noise-induced
effects, namely the resonant activation (RA) \cite{doe}-\cite{anna}
and the noise enhanced stability (NES) \cite{man}-\cite{ale}. The
NES effect increases the average lifetime of the metastable state,
while the RA phenomenon minimizes this lifetime.

The purpose of this work is to find the optimal range of parameters
in which the positive role of resonant activation phenomenon, with
respect to the cancer extinction, prevails over the negative role of
NES, which enhances the stability of the tumoral state. The following
section explains more extensively the concepts of resonant
activation and noise-enhanced stability and their role in the
analysis of cancer growth kinetics. In the next section we report
the appearance of RA and NES phenomena in this model and we discuss
the statistics of first passage times in this regime. In the same
section the role of different potential profiles on the RA and NES
effects, with piece-wise linear and MM potentials with reflecting
boundary is analyzed. Finally we draw the conclusions.

%
%

\section{The Model System}

%
%

\subsection{Mean first passage time analysis}
Our starting point is the following Langevin equation of an
overdamped Brownian particle moving in a random fluctuating
potential barrier
\begin{eqnarray}
\frac{dx}{dt} & =& -\frac{dV(x,t)}{dx}+\xi(t), \nonumber \\
  \nonumber \\
    V(x,t) &=& U(x) + G(x) \eta(t).
 \label{eq:lang2}
\end{eqnarray}

\noindent Here $\xi(t)$ is a Gaussian process with zero mean and
correlation function $\langle \xi(t)\xi(t')\rangle=\sigma^2\delta(t-t')$.
The potential $V(x,t)$ is the sum of two terms: the fixed potential
$U(x)$ and the randomly switching term $G(x) \eta(t)$, where
$\eta(t)$ stands for a Markovian dichotomous noise switching between
two levels $\{\Delta^{+},\Delta^{-}\}$, where $\Delta^{+}=\Delta$ and $\Delta^{-}=-\Delta$.
 The probability of switching $(\pm\Delta \rightarrow \mp\Delta)$ per time unit, or the mean frequency of switching, is labelled $\nu = 1/(2 \tau)$, and the autocorrelation function of the dichotomous noise yields
$$\langle(\eta(t)-\langle\eta\rangle)(\eta(t')-
\langle\eta\rangle)\rangle=\frac{\left(\Delta^+-\Delta^-
\right)^2}{4}e^{-|t-t'|/{\tau}}.$$

\noindent The two noise sources are statistically independent,
\textit{i.e.} $\langle\xi(t)\eta(s)\rangle=0$. The potential
$V(x,t)$ therefore flips at random time between two configurations
\begin{equation}
 U^{\pm}(x) = U(x) + G(x)\Delta^{\pm}.
 \label{pot}
 \end{equation}

\noindent The corresponding Fokker-Planck equations which describe
the evolution of probability density of finding the state variable
 in a ``position'' $x$ at time $t$ are
\begin{eqnarray}
\partial_t {p}(x,\Delta^\pm,t)& =&  \partial_x  \left[\frac{dU^\pm(x)}
{dx}+\frac{1}{2}\sigma^2\partial_x  \right]p(x,\Delta^\pm,t) \nonumber \\
  & -& \frac{1}{2\tau} p(x,\Delta^\pm,t)+\frac{1}{2\tau} p(x,\Delta^\mp,t).
\label{eq:schmidr}
\end{eqnarray}

\noindent In the above equations time has dimension of
$[length]^2/energy$ due to a friction constant that has been
``absorbed'' in a time variable. With the initial condition
\begin{equation}
p(U\pm,x_s,t)|_{t=0}=\frac{1}{2}\delta(x-x_s),
\end{equation}

\noindent from Eqs.\ (\ref{eq:schmidr}) we get the equations for the
mean first passage times (MFPTs):
``	````				`																										`										
\begin{eqnarray}
\label{eq:mfpt bn}
-1 &=& -\frac{T^+(x)}{\tau}
+\frac{T^-(x)}{\tau}- 2\frac{dU^+(x)}{dx}\frac{dT^+(x)}{dx}
+\sigma^2\frac{d^2 T^+(x)}{dx^2}\nonumber \\
-1 &=& \frac{T^+(x)}{\tau}-\frac{T^-(x)}{\tau}-
2\frac{dU^-(x)}{dx}\frac{d T^-(x)}{dx} +\sigma^2\frac{d^2
T^-(x)}{dx^2},
 \end{eqnarray}
where $T^+(x)$ and $T^-(x)$ denote MFPT for $U^+(x)$ and $U^-(x)$,
respectively. The overall MFPT  for the system is derived from the above set of equations $T(x)= T^+(x)+T^-(x)$
by assuming a reflecting boundary condition at $x=a$ and an absorbing
boundary at $x=b$:
\begin{eqnarray}
\frac{dT^{\pm}(x)}{dx}|_{x=a}=0,\nonumber \\
T^{\pm}(x)|_{x=b}=0,
 \end{eqnarray}
Analytical solutions of Eqs.~(\ref{eq:mfpt bn}) can be
expressed in a compact form only for
piece-wise linear or piece-wise constant potentials \cite{doe,bie}.
More complex cases require either use of approximation schemes
\cite{andrey,boguna,rei}, perturbative approach \cite{iwa}, or
direct numerical evaluation methods \cite{marchi,ms,dybiec}.

%
%
\subsection{Resonant activation}\label{subsec:ra}

The signature of the resonant activation phenomenon is the
occurrence of a minimum of the MFPT as a function of the mean
frequency of the switching $\nu$ of the multiplicative dichotomous
noise. As a consequence the resonant activation effect minimizes
 the average lifetime of a population in the metastable state.
Let us assume that the Brownian particle is behind the potential
barrier, in a neighborhood of the metastable state. When the barrier
fluctuations are very fast the particle ``sees" the average
potential between the higher and lower configurations. The MFPT will
tend to a constant value corresponding to the average static
potential
\begin{equation}
\lim_{\tau\rightarrow 0}MFPT(U^+,U^-,\tau)=
MFPT\left(\frac{U^+}{2}+\frac{U^-}{2}\right).
\label{eq:MFPT_zero}
\end{equation}
If the barrier fluctuations are slower than the actual escape rate
($\tau$ large), the particle will escape before any barrier flip
occurs. Therefore, the mean first passage time also tends to a
constant, which now will be an average of the escape times for the
higher and lower configurations of the potential
 \begin{equation}
\lim_{\tau\rightarrow\infty}MFPT(U^+,U^-,\tau)=
\frac{1}{2}\left[MFPT(U^+)+MFPT(U^-)\right]. \label{eq:MFPT_inf}
\end{equation}
If the potential barrier is high enough or the additive noise is
weak enough, the mean first passage time in a static potential can
be approximated by the inverse of the Kramers escape rate, which
increases exponentially with $\Delta U(x)/\sigma^2$, where $\Delta
U(x)$ is the height of the barrier. Such an exponential dependence
guarantees the existence of a minimum of MFPT $(\tau)$. Its value at
intermediate $\tau$ is lower than both asymptotic mean first passage
times, for very large or very small $\tau$~\cite{bdka}.
Specifically, the MFPT at very low mean switching frequency
(Eq.~(\ref{eq:MFPT_inf})) will be higher than the MFPT at the high
frequency limit (Eq.~(\ref{eq:MFPT_zero})). In the middle frequency
regime the effective escape rate over the fluctuating barrier is the
average of the escape rates

\begin{equation}
K_{eff} = \frac{1}{2}(K_+ +K_-),
\end{equation}
where $K_+ = 1/MFPT(U^+)$ and $K_- = 1/MFPT(U^-)$, with $K_-\ll
K_+$. Because of the exponential dependence, the MFPT in this
intermediate frequency regime will be smaller than both
$MFPT_{\tau\rightarrow \infty}$ and $MFPT_{\tau\rightarrow 0}$.

%
%
\subsection{Noise-enhanced stability}

The signature of the noise-enhanced stability effect is the
occurrence of a maximum of the MFPT as a function of the additive
noise intensity, for a fixed mean frequency $\nu$ of the
multiplicative noise. As a consequence the NES effect, differently
from the RA phenomenon, maximizes the average lifetime of the
population in the metastable state. The nonmonotonic behavior of the
mean escape time as a function of the additive noise intensity
$\sigma$ depends on the potential profile parameters, on the
parameters of the multiplicative dichotomous noise and also on the
initial position of the Brownian particle~\cite{as}-\cite{dub}.
Noise enhances the stability of the metastable state with different
peculiarities related to different dynamical regimes: the average
lifetime can greatly increase when the noise intensity is very low
with respect to the height of the barrier and the initial positions
of the Brownian particles are in the ``divergent" dynamical regime
~\cite{ale,bork}. When the particle starts from initial positions within
the potential well, at small values of $\sigma$ with respect the
height of the potential barrier and for a fixed potential, it will
follow the monotonic behavior of the Kramers formula, as a function
of the noise intensity. The mean escape time will then tend to
infinity for $\sigma \rightarrow 0$. If the particle starts from
outside the well, at small $\sigma$ it will at once run down from
its starting region towards the absorbing barrier and its mean
escape time will be approximately equal to the escape time of a
deterministic particle. At high noise intensities, the mean escape
time decreases monotonically, regardless of initial positions. At
intermediate noise intensities, a particle starting from the outer
slope may sometimes be trapped into the well. Such events, although
rare, can significantly increase the mean escape time because a
trapped particle stays then in the well for a relatively long time.

%

\subsection{Population model}

The kinetics of our biological system is described by the equation
\begin{equation}
 \frac{dx}{dt}=(1-\theta x)x - \beta\frac{x}{x+1},
\label{eq:system}
\end{equation}
where $x(t)$ is the concentration of the cancer cells. The profile
of the corresponding quasi-potential (Eq.~\ref{eq:pot}) presents a
double well with one of the minima at $ x=0 $. The region for $x>0$
can show either a monotonic behavior or a local minimum, depending
on the values of parameters $\theta$ and $\beta$. In the present
investigation we used only parameters able to give a local minimum
of the Michaelis-Menten potential for $x>0$: $\theta=0.1$ and
$\beta= 3$ (see Fig.~\ref{fig:pot}). For $x \rightarrow \infty$ the
potential shows a strong cubic repulsion. Our biological system is
an open system whose random interactions with the environment and
the random fluctuations of the immune system are described by an
additive noise term $\sigma\xi(t)$ in Eq.~(\ref{eq:system}), and a
dichotomous Markovian noise $\eta(t)$, of amplitude $\Delta$ and
mean correlation time $\tau$ in the $\beta$ parameter. As a
consequence the stochastic Michaelis-Menten (MM) potential switches
between two conformational states $U^\pm(x)$
 \begin{equation}
U^\pm(x)=-\frac{x^2}{2}+\frac{\theta x^3}{3}+(\beta \pm \Delta)( x -
\ln(x+1)),
\end{equation}
and the Langevin equations for the system are
 \begin{eqnarray}
\dot{x} &=& -\frac{dU^\pm(x)}{dx} + \xi(t) \nonumber \\
&=& x(1-\theta x) - (\beta \pm \Delta) \frac{x}{x+1} + 
\xi(t).
 \label{eq:langevin}
 \end{eqnarray}
The process of population growth and decay can be described as a
motion of a fictitious particle in the switching potential between
two configurations $U^+(x)$ and $U^-(x)$. For negligible additive
noise and small concentration of cancerous cells, this model
resembles a standard Verhulst equation with perturbing
multiplicative dichotomous noise, which exhibits a complex scenario
of noise-induced transitions, observable in a pattern of the
stationary probability density~\cite{Lefever}. Here, we will address
kinetic properties of this model by studying the mean first passage
time cite{bork} between high and low population states in the
system. We will study how the two different sources of noise as well
as the position of the starting point $x_{\mathrm{in}}$ influence
the mean first passage time. We put the absorbing boundary at $x=0$
and the reflecting one at $x=\infty$. The event of passing through
the absorbing boundary
 is equivalent to a total extinction of cancer.


\section{Results}

\subsection{Michaelis-Menten and piece-wise linear potentials with absorbing boundary}

Here we analyze the role of different potential profiles with a
reflecting barrier placed on the minimum of the fixed MM potential,
that is in absence of dichotomous noise. We consider two cases: (i)
MM potential, and (ii) piece-wise linear potential. Both cases are
approximations of the effective MM potential (Fig.~\ref{fig:pot}.
Particularly the piece-wise linear potential is a good theoretical
model, which enable to obtain closed analytical expressions of the
MFPT (see Refs.~\cite{doe,bie}). In the following 
Fig.~\ref{fig:MMapprox} we report the results of our simulations on
both cases and compare the behaviors of MFPT for NES and RA
phenomena with those obtained using the exact MM potential profile
(see Fig.~\ref{fig:pot}). The NES effect, for a fixed value of the
correlation time of the multiplicative noise and an unstable initial
position, is shown in Fig.~\ref{fig:MMapprox}a. The reflecting
barrier placed in the proximity of the metastable state pushes the particle away
from it and therefore decreases the MFPT for all values
of the noise intensity, producing an overall reduction of the NES
effect. The Brownian particle experiences only two different slopes
in the piece-wise linear potential, while more slopes are
experienced, and therefore more time spent around the metastable
state, in the MM potential~\cite{as}. This explain why the piece-wise
linear potential is the worst approximation. For noise intensity
values greater than the height of the potential barrier, the
different shape of the potential profile becomes irrelevant and the
decreasing of MFPT values is due to the reflecting boundary only. In
Fig.~\ref{fig:MMapprox}b is reported the MFPT as a function of the
correlation time of the multiplicative noise, at a fixed value of
the noise intensity, for the same different potential profiles. The
effect of the reflecting boundary is again to decrease all the MFPT
values in the MM potential and therefore to reduce the absolute
values of the RA phenomenon. The minimum of MFPT is of course more
pronounced in the exact MM potential and the piece-wise linear
potential is again the worst approximation. Both effects are robust
enough to be observed in different potential profiles.
%
%
\begin{figure}[t]
\begin{center}
\epsfig{figure=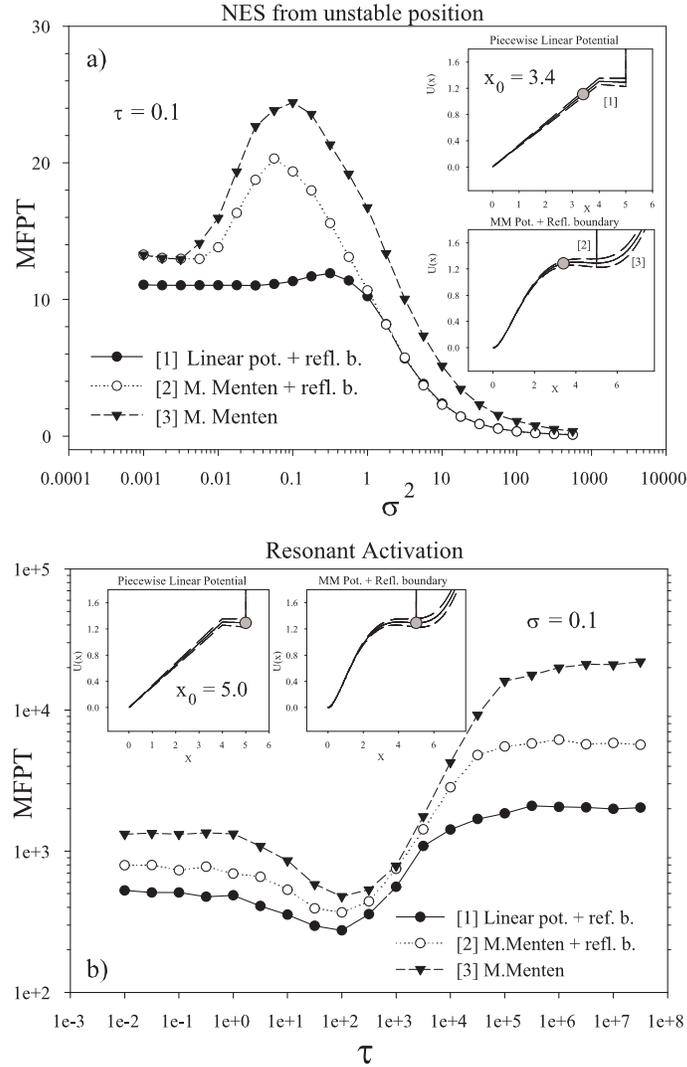, width=9.2cm}
\end{center}
\caption{\label{fig:MMapprox} (a) MFPT as a function of the additive
noise intensity $\sigma^2$ for three potential profiles: [1]
piece-wise linear with a reflecting boundary, with slopes $k_1 =
+0.32$ and $k_2 = -0.025$ in the average position; [2]
Michaelis-Menten (MM) with a reflecting boundary; [3] exact MM
potential. The unstable initial position is at $x_0 = 3.4$ and $\tau
= 0.1$. (b) MFPT as a function of the correlation time of the
multiplicative noise. The initial position is $x_0=0.1$, $\sigma =
0.1$ and the number of realizations is $N = 10^4$. All the other
potential parameters are the same as in Fig.~\ref{fig:pot}.}
\end{figure}
%
%

\subsection{Coexistence of noise-enhanced stability and resonant activation}

In this section we look for the coexistence region of RA and NES
effects. We calculate therefore the MFPT for the following initial
positions $x_{\mathrm{in}}=3,3.65,5$. We performed a series of Monte
Carlo simulations of the stochastic process (\ref{eq:langevin}) with
an absorbing boundary at $x=0$, natural reflecting boundary at
$x=+\infty$ and the values of parameters: $\beta=3$, $\theta=0.1$,
$\Delta=\pm 0.02$. The statistics for each MFPT has been taken from
$N=10^4$ (for $x_{\mathrm{in}}=3.65$) and $N=10^3$ (for
$x_{\mathrm{in}}=3$ and $5$) simulation runs. The results reported
here are an extension of those presented in \cite{ochab_pre}. The
results confirm the existence of resonant activation and
noise-enhanced stability phenomena in the studied system. Moreover,
we have shown that in a certain range of parameters both effects can
occur together \cite{dub,ps,eps}.

\begin{figure}[t]
\epsfig{figure=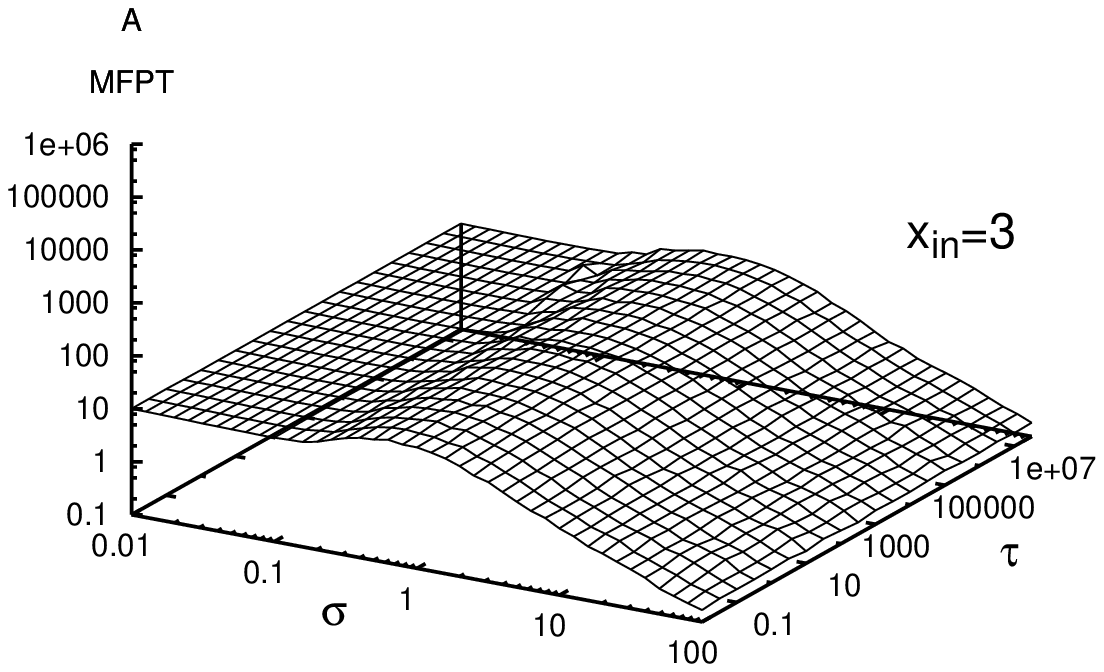, width=6cm} \epsfig{figure=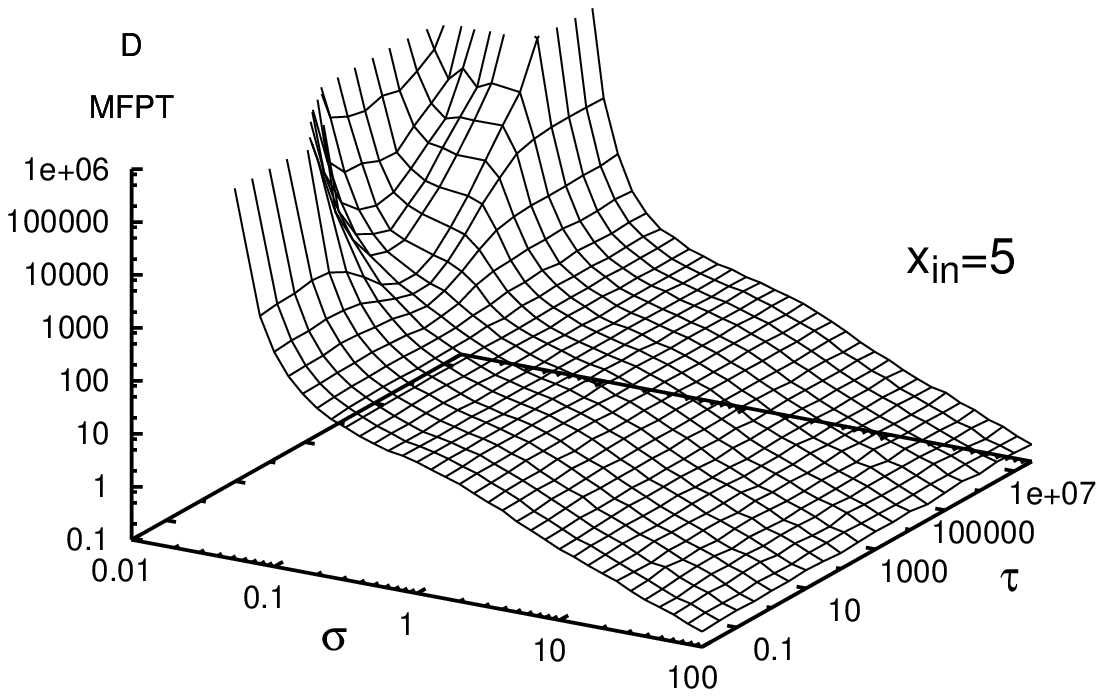,
width=6cm} \caption{\label{fig:3d_3_5} MFPT as a function of the
additive noise intensity $\sigma$ and the correlation time of
multiplicative noise. Initial positions: $x_{\mathrm{in}}=3$ (left
panel), $x_{\mathrm{in}}=5$ (right panel). Parameters: $\beta=3$,
$\theta=0.1$, $\Delta=\pm 0.02$. Number of simulation runs per each
point: $N=10^3$.}
\end{figure}

\begin{figure}[t]
\epsfig{figure=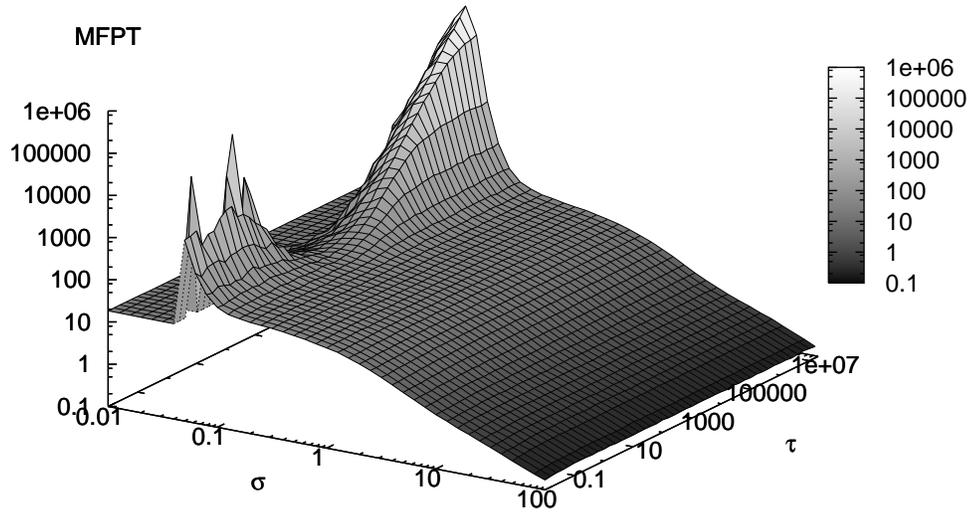, width=13cm} \caption{\label{fig:3d_3.65} MFPT
as a function of the additive noise intensity $\sigma$ and the
correlation time of multiplicative noise. Initial position:
$x_{\mathrm{in}}=3.65$. The parameters are the same as in
Fig.~\ref{fig:3d_3_5}. Number of simulation runs per each point:
$N=10^4$.}
\end{figure}

\begin{figure}[h]
\epsfig{figure=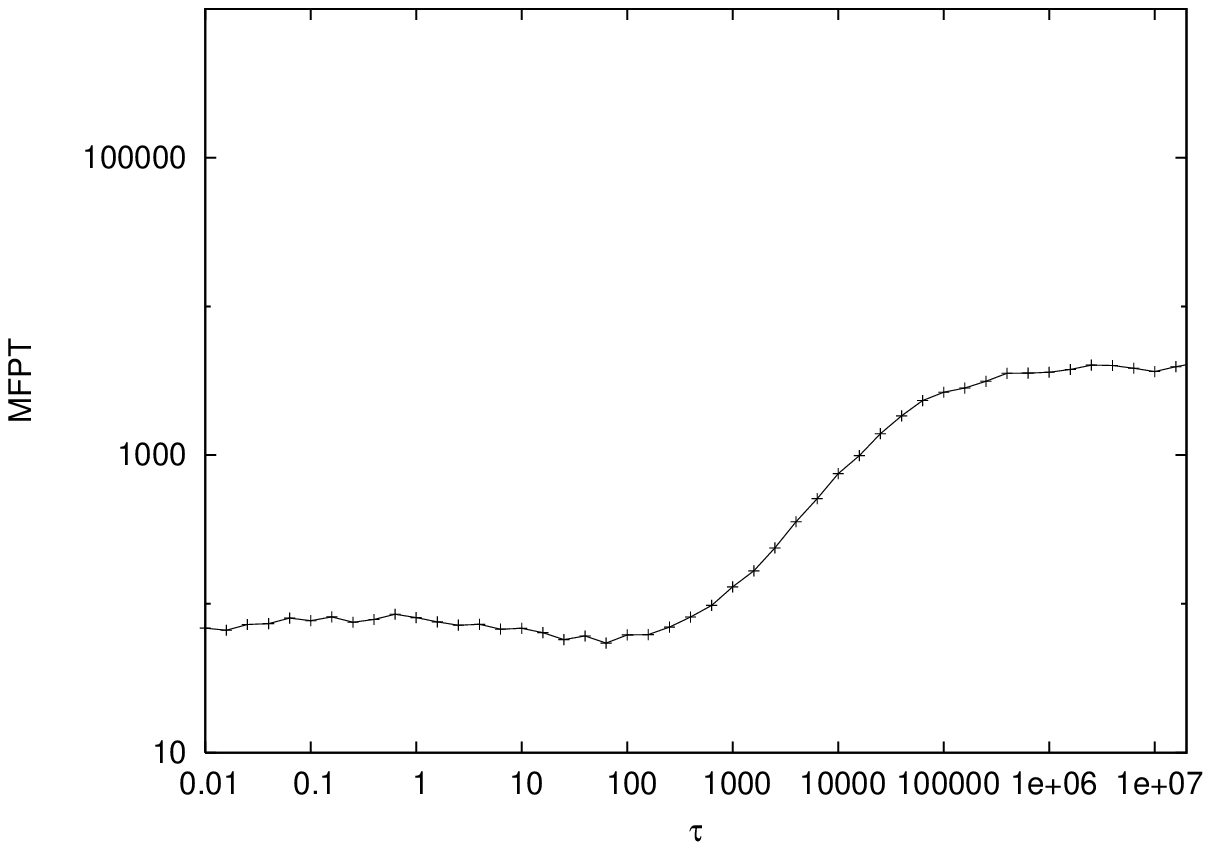, width=6cm}
\epsfig{figure=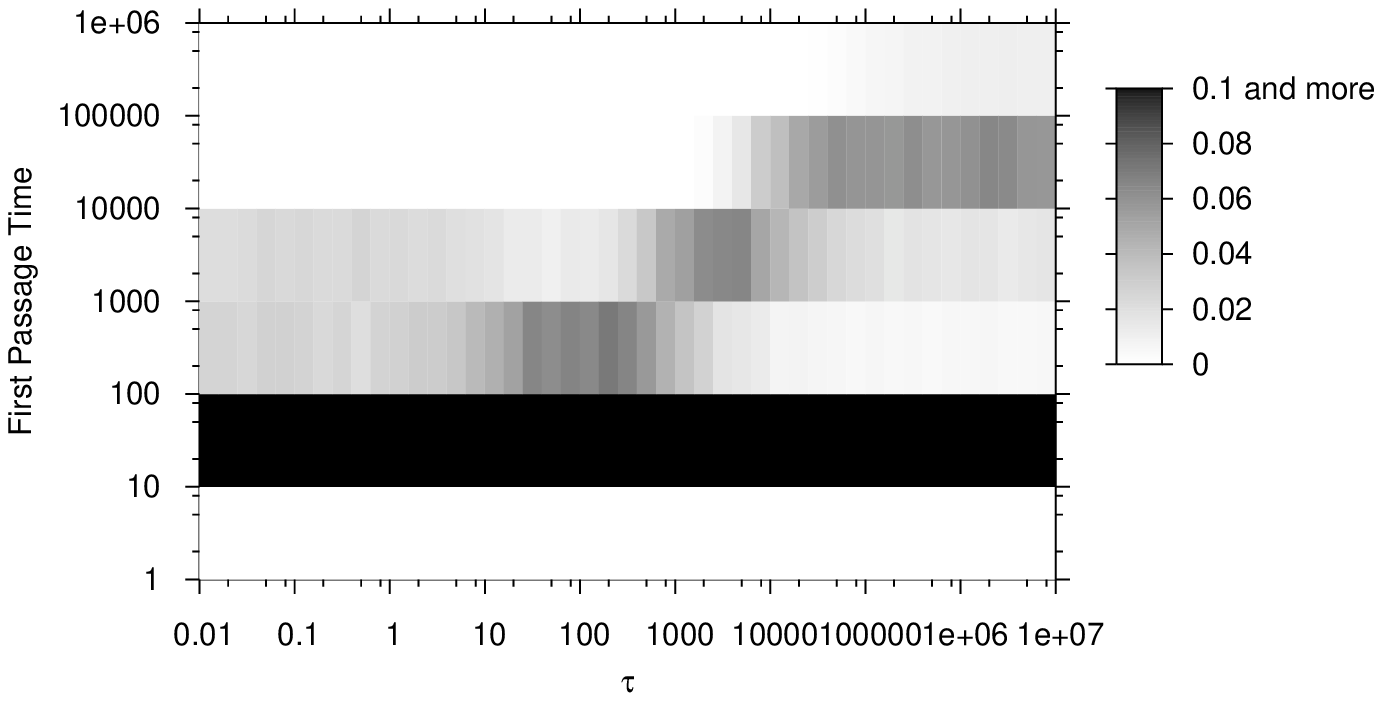, width=7cm}\\
\epsfig{figure=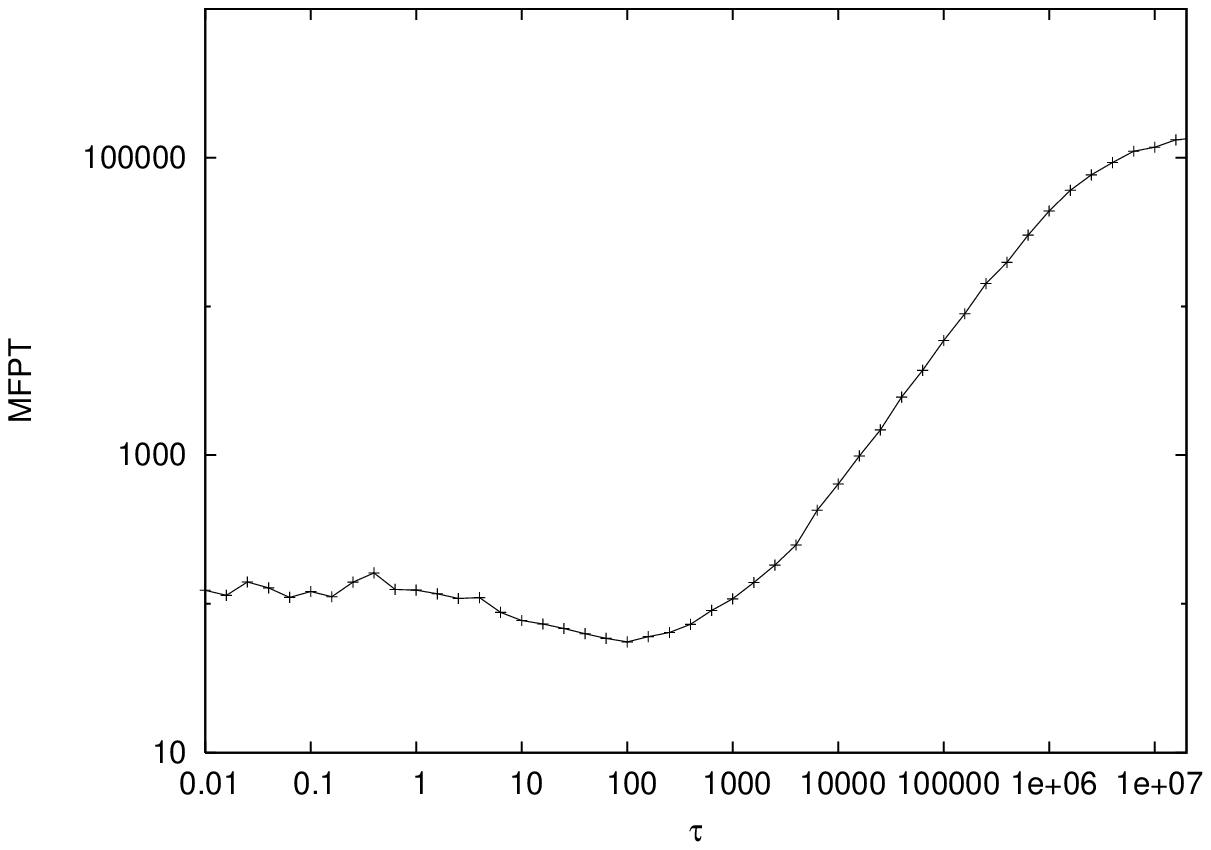, width=6cm}
\epsfig{figure=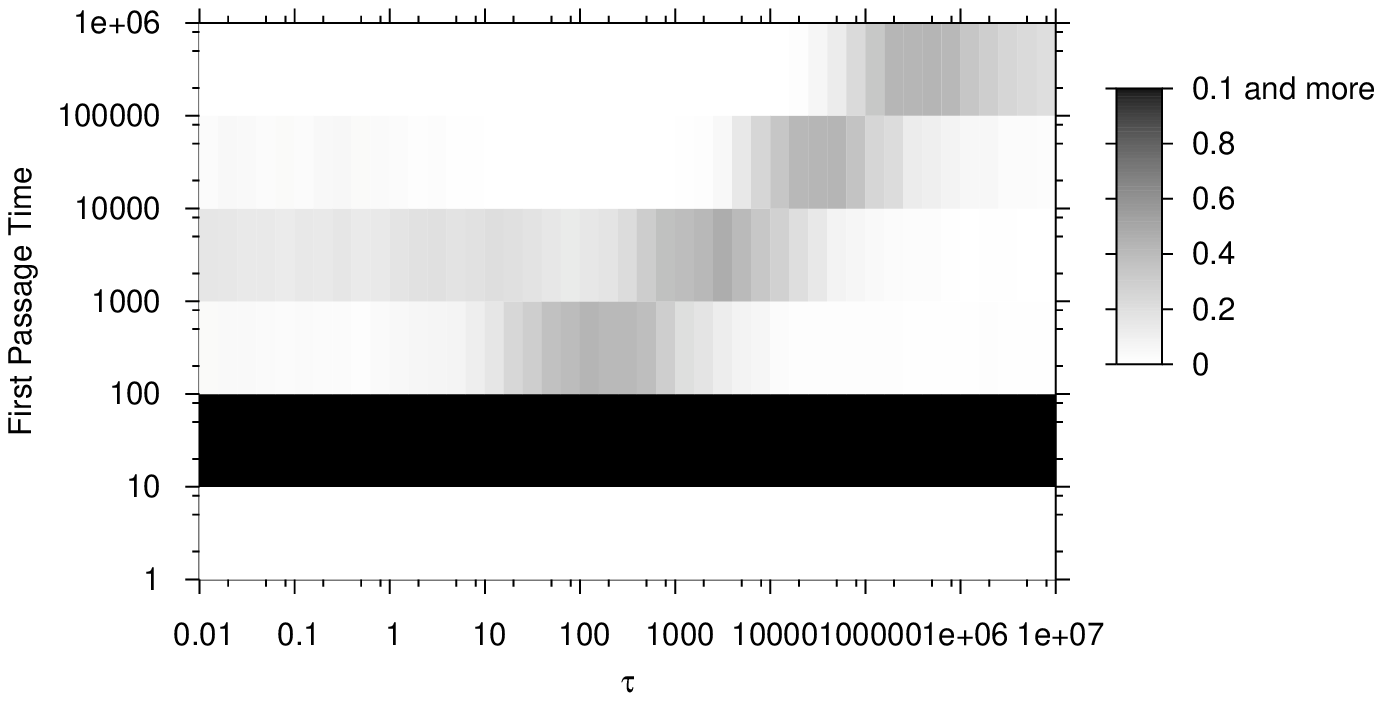, width=7cm}\\
\epsfig{figure=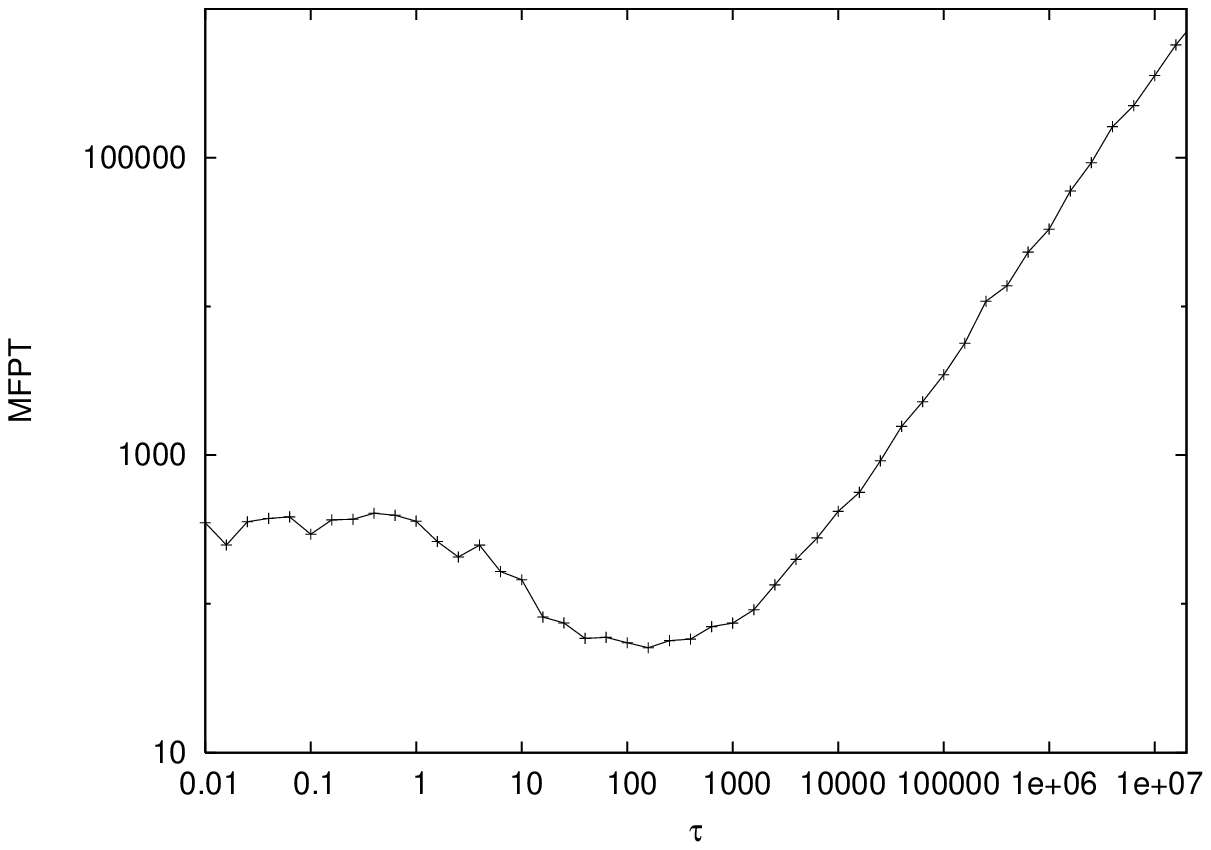, width=6cm}
\epsfig{figure=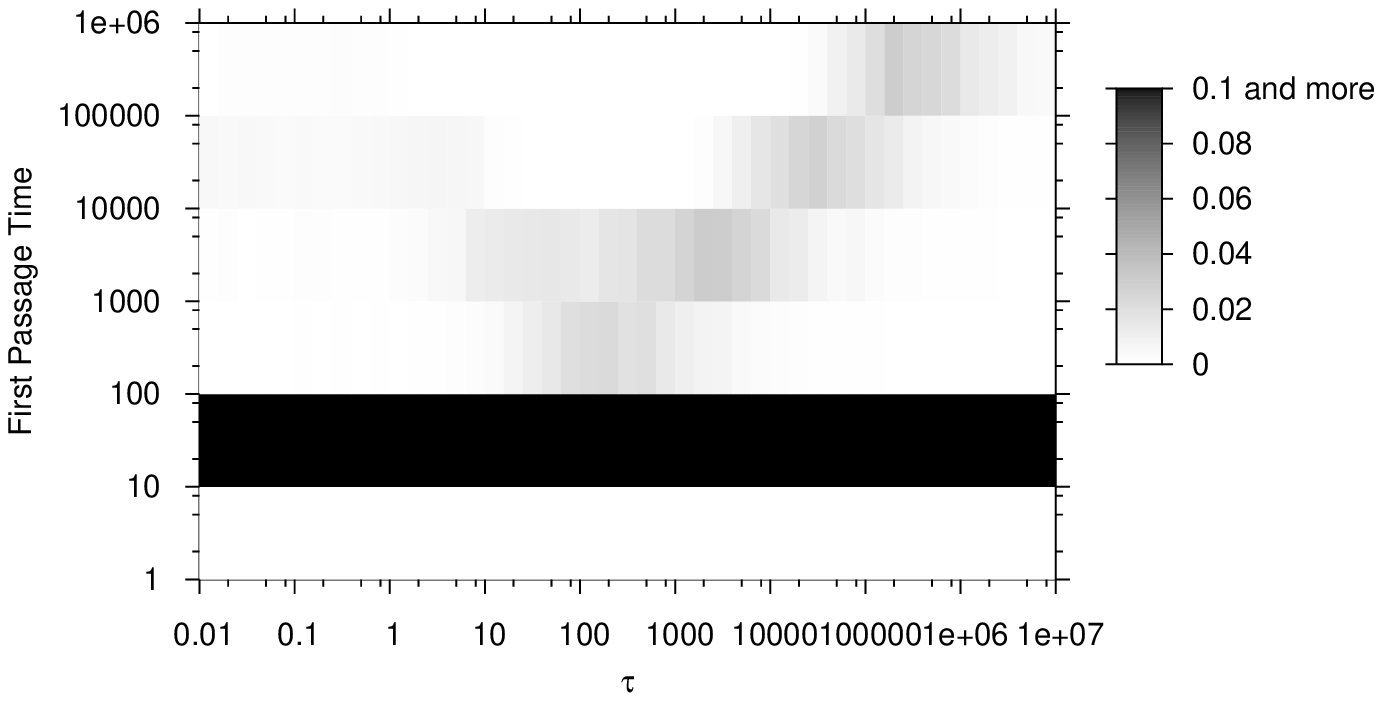, width=7cm}\\
\caption{\label{fig:histogram1} Left panels: MFPT as a function of
$\tau$ for chosen values of $\sigma$ in the noise-enhanced stability
regime. From top to bottom: $\sigma=0.10$, $\sigma=0.08$,
$\sigma=0.06$. Right panels: Corresponding first passage time
histograms. Colors from white to black denote a fraction of all
trajectories. Black color denotes values from 0.1 to 1. Initial
position: $x_{\mathrm{in}}=3.65$. The other parameters are the same
as in Fig.~\ref{fig:3d_3_5}.}
\end{figure}

In Figs.~\ref{fig:3d_3_5} and~\ref{fig:3d_3.65} we present the
combined view of RA and NES effects. The considerations of
subsection~\ref{subsec:ra} are valid not only for trajectories
starting from inside the potential well, but also for arbitrary
initial positions in the potential, if only the particle has a
chance to be trapped behind the potential barrier for some time. If
the additive noise intensity is very large, a particle starting from
the outer slope of the barrier does not ``feel" the barrier at all.
If, in turn, the additive noise is very weak, the particle slides
down the slope in an almost deterministic way. But at intermediate
values of $\sigma$, the particle can (in some, very infrequent,
realizations of the stochastic process) be trapped behind the
barrier and, in case of such a rare event, its escape time changes
non-monotonically as a function of $\tau$, in a way described in
subsection~\ref{subsec:ra}. This effect of coexistence of
noise-enhanced stability and resonant activation effects can be
observed in Fig.~\ref{fig:3d_3.65}. When the trajectory starts on
the outer slope of the potential, only small fraction of the
realizations are trapped into the well, and thus the RA effect is
very small because only few trajectories will contribute to it. In
order to obtain a well-visible RA (a large enough number of trapped
trajectories)  we have to choose the initial position sufficiently
close to the top of $U^-(x)$. We have chosen $x_{\mathrm{in}}=3.65$.
For this initial position we observe a co-occurence of NES and RA.
Since both effects act here in an opposite way, there exists a
regime of $\sigma$ and $\tau$ parameters where noise-enhanced
stability is strongly reduced by resonant activation. This region is
well visible in Fig.~\ref{fig:3d_3.65} for values of parameters:
$\sigma \approx 0.06$, $\tau \approx 100$. In
Fig.~\ref{fig:histogram1} we present cross-sections of
Fig.~\ref{fig:3d_3.65} at chosen values of $\sigma$ and histograms
of the first passage times corresponding to each point on the
graphs. Most trajectories (more than $90\%$ of all) are those which
do not fall into the potential well. The trajectories trapped are
minor contibutions: their fraction (of order of $0.1$, $0.01$ and even
$0.001$) depends on the barrier switching rate and on the additive
noise strength. At $0.01<\sigma < 1$, the first passage times of
particles, which had not been trapped into the well, are of order of
$10$. The first passage times of the trajectories trapped behind the
potential barrier depend of course on $\tau$: In the region where
the resonant activation is better visible, for small $\tau$ the
first passage times are of order of $10^3$, for large $\tau$ they
are of order of $10^6$ and more, and for the intermediate values of
$\tau$ the first passage time drops down to the order of $10^2$.

%


\section{Conclusions}

We studied a Langevin equation derived from the phenomenological
Michaelis-Menten scheme for catalysis accompanying a spontaneous
replication of molecules. It contains an additive noise term
(Gaussian white noise) and a multiplicative noisy driving
(dichotomous noise) in the term responsible for inhibition of
population growth. The models of that type are widely used in biophysical description  
of cell-mediated immune surveillance against cancer \cite{Prigogine,GARAY,lefever,vladar,steel}, analysis of cell survival including repair kinetics after exposure to ionizing radiation \cite{sontag} or delineation of factors responsible for the efficient action of biological regulatory networks \cite{Paulsson,rao,pomerening,alper}.
 We examined how
the two different sources of noise influence the population's
extinction time, identified with the mean first passage time of the
system over the zero population state. We observe appearance of two noise-induced
resonant phenomena in the system: Mean first passage time displays here a non-monotonic behavior which depends on the characteristic
parameters of the noises present in the system, namely the additive
noise intensity $\sigma$ and the correlation time $\tau$ of the
multiplicative noise. The occurence of a minimum in MFPT (resonant
activation) is related to the $\tau$ parameter, whereas the
emergence of the maximum in MFPT (noise-enhanced stability) is
connected with $\sigma$. We report the evidence for co-occurrence of
resonant activation and noise-enhanced stability in the same regime
of noise parameters. We observe that the strong stability
enhancement of the metastable state comes from very few
contributions from the trajectories trapped in it. Their first
passage times are  several orders of magnitude longer than the first
passage times of the trajectories which approached the absorbing
boundary without being trapped in the metastable state. We analyzed
the role of different potential profiles, with a reflecting barrier,
on RA and NES phenomena. We found a very interesting, from the point of view of growth kinetics of cancer cells, coexistence
region of both effects. In this region the NES effect, which enhances the
stability of the tumoral state, is strongly reduced by the RA
effect, which enhances the cancer extinction. In other words, an asymptotic regression to the zero tumor size may be induced by controlling the noise affecting hyperbolic inhibition of a spontaneous proliferation of cells.  

\section{Acknowledgements}

This work was supported by the European Science Foundation STOCHDYN
grant, MIUR and INFM-CNR.
The paper is dedicated to Professor Pater Talkner on the occasion of his 60th birthday anniversary.

%

%

\end{document}